\documentclass[11pt,a4paper]{article}
\usepackage{amsmath}
\usepackage[mathcal]{eucal}
\usepackage{latexsym}
\usepackage{amssymb}
\begin{titlepage} 
\title{Nonlinear gravitons in 4-D general relativity by expansion about the Kodama state}
\author{Eyo Eyo Ita III}
\medskip
\input amssym.def
\input amssym.tex

\def \in{\indent}

\begin{document}
\maketitle
\bigskip
\centerline{Department of Applied Mathematics and Theoretical Physics} 
\smallskip
\centerline{Centre for Mathematical Sciences, University of Cambridge, Wilberforce Road}
\smallskip
\centerline{Cambridge CB3 0WA, United Kingdom}
\smallskip
\centerline{eei20@cam.ac.uk} 

\bigskip

\begin{abstract}
In this paper we provide a possible realization of Penrose's idea of nonlinear gravitons using a new description of nonmetric general relativity.  In the addressal of issues surrounding the normalizability of the Kodama state and its reliability as a ground state for gravity, we expand the theory in fluctuations about the Kodama state.  This produces a theory of complex gravity with a well-defined Hilbert space structure, whose quantization we carry out both at the linearized level and in the full nonlinear theory.  The results of this paper demonstrate the preservation of the physical degrees of freedom of the full nonlinear theory under linearization, as well provide a Hilbert space of states of the former annihilated by the quantum Hamiltonian constraint.
\end{abstract}
\end{titlepage}

\section{Introduction}

In \cite{PENROSE} Roger Penrose takes issue with the standard view of the graviton as a weak-field perturbation of a background spacetime.  He proposes the idea that each graviton should carry its measure of curvature, corresponding to a solution of the full nonlinear Einstein equations.  The issue of the graviton is of supreme importance, particularly when one wishes to construct a quantum theory of gravity.  The Penrose approach in \cite{PENROSE} presents the concept of left-handed and right-handed gravitons, which leads to a twistor theory naturally adapted to the description of complex spacetimes.  In this paper we will demonstrate a realization of Penrose's idea using a nonmetric description of complex GR which is different from twistor theory, and is derivable from the Ashtekar formulation.\par
\indent
In the complex Ashtekar theory of gravity the basic phase space variables are a self-dual $SO(3,C)$ connection and a densitized triad $\Omega_{Ash}=(A^a_i,\widetilde{\sigma}^i_a)$.  The action in 3+1 form is a canonical one form minus a linear combination of first class constraints smeared by auxilliary fields \cite{ASH1}, \cite{ASH2}, \cite{ASH3}

\begin{eqnarray}
\label{THEACTION}
I_{Ash}=\int{dt}\int_{\Sigma}d^3x\Bigl[\widetilde{\sigma}^i_a\dot{A}^a_i+A^a_0G_a-N^iH_i-i\underline{N}H\Bigr].
\end{eqnarray}

\noindent
The fields $N^i$, $A^a_0$ and $\underline{N}=N(\hbox{det}\widetilde{\sigma})^{-1/2}$ are respectively the shift vector, temporal component of a 4-D self-dual connection $A^a_{\mu}$, and the lapse density function.  The initial value constraints are the diffeomorphism constraint $H_i$, given by

\begin{eqnarray}
\label{THEACTION1}
H_i=\epsilon_{ijk}\widetilde{\sigma}^j_aB^k_a=0,
\end{eqnarray}

\noindent
the Gauss' law constraint $G_a$ which is given by

\begin{eqnarray}
\label{THEACTION2}
G_a=D_i\widetilde{\sigma}^i_a=0,
\end{eqnarray}

\noindent
and the Hamiltonian constraint $H$ by 

\begin{eqnarray}
\label{THEACTION3}
H=\epsilon_{ijk}\epsilon^{abc}\widetilde{\sigma}^i_a\widetilde{\sigma}^j_b\Bigl(B^k_c+{\Lambda \over 3}\widetilde{\sigma}^k_c\Bigr)=0
\end{eqnarray}  

\noindent
where $\Lambda$ is the cosmological constant.  There is a nontrivial solution to the system (\ref{THEACTION1}), (\ref{THEACTION2}) and (\ref{THEACTION3}) given by $\widetilde{\sigma}^i_a=-{3 \over \Lambda}B^i_a$ which enables one to construct a Hamilton--Jacobi functional $I_{CS}[A]$, namely the Chern--Simons functional of the spatial connection $A^a_i$.  The exponentiation of this functional yields the Kodama state, which was first discovered by Hideo Kodama \cite{KOD}

\begin{eqnarray}
\label{THEACTION4}
\boldsymbol{\psi}_{Kod}[A]=e^{-3(\hbar{G}\Lambda)^{-1}I_{CS}[A]}.
\end{eqnarray}

\noindent
Equation (\ref{THEACTION4}) exactly solves the classical constraints and also the quantum constraints of GR for a particular operator ordering \cite{POSLAMB}.  There are some objections to the use of $\boldsymbol{\psi}_{Kod}$ as a ground state for gravity, by analogy to the pathologies of the Chern--Simons functional when seen in the purely Yang--Mills context.  These pathologies include nonnormalizability and nonunitarity, as well as the lack of a reliable Hilbert space structure \cite{WITTEN1}.\par
\indent
We will address these objections by showing that there exists a well-defined theory of fluctuations about $\boldsymbol{\psi}_{Kod}$, seen as the ground state for some gravitational system.  In this paper we will show demonstrate that the fluctuations take on the interpretation of gravitons, and we will quantize these fluctuations and show that they admit a genuine Hilbert space.  This task has been carried out to some extent at the 
linearized level in \cite{LINKO} in the Ashtekar variables.  In the present paper we will extend the demonstration to Lorentzian signature spacetimes, using a new set of phase space variables $\Omega_{Inst}=(\Psi_{ae},A^a_i)$ which will be defined later.  We will carry out the demonstration both at the linearized level and for the full nonlinear theory.  In this paper we will not address reality conditions, which is treated elsewhere.\par
\indent

\subsection{Organization of this paper}

The organization of this paper is as follows.  After transforming from the Ashtekar phase space $\Omega_{Ash}=(\widetilde{\sigma}^i_a,A^a_i)$ into the new phase space variables variables $\Omega_{Inst}=(\Psi_{ae},A^a_i)$ we expand the starting action including the initial value constraints about the action associated with the Kodama state $\boldsymbol{\psi}_{Kod}$.  Prior to embarking upon the full theory we first demonstrate the expected features of the graviton in the linearized approximation in Part I.  Section 3 performs a linearization about $\boldsymbol{\psi}_{Kod}$ on $\Omega_{Inst}$, producing the massless spin two polarizations in this limit.  We then perform a quantization 
on $\Omega_{Inst}$, demonstrating the existence of a Hilbert space structure at the linearized level.  In Part Two we redo the previous exercises, now with respect to the full nonlinear theory.  First we put in place the requisite canonical structure for quantization, which entails the implementation of the kinematic constraints at the level of the starting action.  Then we introduce the auxilliary Hilbert space and use it as a basis for construction of wavefunctions annihilated by the Hamiltonian constraint of the full theory.  A time variable $T$ on configuration space emerges similarly to the case in the linearized theory, and the wavefunction evolves with respect to this time in the full theory.  The well-definedness of the quantization is linked to the convergence of solutions of the full Hamiltonian constraint with respect to time variable $T$, which we prove in this section.\par
\indent
To proceed from the Ashtekar phase space $\Omega_{Ash}$ into the new phase space $\Omega_{Inst}$, let us first make a substitution called the CDJ Ansatz

\begin{eqnarray}
\label{THEACTION41}
\widetilde{\sigma}^i_a=\Psi_{ae}B^i_e,
\end{eqnarray}

\noindent
where $\Psi_{ae}\in{SO}(3,C)\otimes{SO}(3,C)$ is the CDJ matrix.\footnote{This Ansatz is attributable to Capovilla, Dell and Jacobson, was used in \cite{CAP} to write down a general solution to the Hamiltonian and the diffeomorphism constraints, which are algebraic in nature.  Equation (\ref{THEACTION4}) holds as long as $(\hbox{det}B)\neq{0}$ and $(\hbox{det}\Psi)\neq{0}$.  Since $\widetilde{\sigma}^i_a$ is dimensionless and $[A^a_i]=1$, then it 
follows that $[\Psi_{ae}]=-2$.}  The action (\ref{THEACTION}) then becomes

\begin{eqnarray}
\label{THEACTION5}
I_{Inst}=\int{dt}\int_{\Sigma}d^3x\Bigl[\Psi_{ae}B^i_e\dot{A}^a_i+A^a_0G_a-N^iH_i-iNH\Bigr],
\end{eqnarray}

\noindent
where the corresponding constraints are given by

\begin{eqnarray}
\label{THEACTION6}
H_i=\epsilon_{ijk}B^j_aB^k_e\Psi_{ae}
\end{eqnarray}

\noindent
for the diffeomorphism constraint,

\begin{eqnarray}
\label{THEACTION7}
G_a=B^i_eD_i\Psi_{ae}+C_{be}\bigl(f_{abf}\delta_{ge}+f_{ebg}\delta_{af}\bigr)\Psi_{fg}\equiv\textbf{w}_e\{\Psi_{ae}\}
\end{eqnarray}

\noindent
for the Gauss' law constraint, and

\begin{eqnarray}
\label{THEACTION8}
H=(\hbox{det}B)\bigl({1 \over 2}Var\Psi+\Lambda\hbox{det}\Psi\bigr)=0
\end{eqnarray}

\noindent
for the Hamiltonian constraint where $Var\Psi=(\hbox{tr}\Psi)^2-\hbox{tr}\Psi^2$.  In (\ref{THEACTION7}) we have defined a magnectic helicity density matrix $C_{be}\equiv{A}^b_iB^i_e$ and we have made use of the definition of the covariant derivative of $\Psi_{ae}$, seen as a second-rank $SO(3,C)$ tensor.  In the language of the new phase space variables $\Omega_{Inst}=(\Psi_{ae},A^a_i)$, the Kodama state 
corresponds to the solution $\Psi_{ae}=-{3 \over \Lambda}\delta_{ae}$.\par
\indent
The variation of the canonical one form corresponding to (\ref{THEACTION5}) is given by

\begin{eqnarray}
\label{THEACTION9}
\delta\boldsymbol{\theta}_{Inst}=\delta\Bigl(\int_{\Sigma}d^3x\Psi_{ae}B^i_e\delta{A}^a_i\Bigr)\nonumber\\
=\int_{\Sigma}d^3x\Bigl[B^i_e{\delta\Psi_{ae}}\wedge{\delta{A}^a_i}+\Psi_{ae}\epsilon^{ijk}{(D_j\delta{A}^e_k)}\wedge{\delta{A}^a_i}\Bigr],
\end{eqnarray}

\noindent
which owing to the second term is not a symplectic two form of canonical form.  This features poses an obstruction to the quantization of the theory, which we will show becomes eliminated when one restricts oneself to the linearized level.  In the nonlinear case a transformation from $\Omega_{Inst}$ into new densitized variables is required on the kinematic phase space $\Omega_{Kin}$.

\section{Expansion of the classical constraints relative to the pure Kodama state}

\noindent
We now return to the starting theory defined on $\Omega_{Inst}$ and expand the initial value constraints in fluctuations about the pure Kodama state $\boldsymbol{\psi}_{Kod}$.  We will use the Ansatz

\begin{equation}
\label{KODD}
\Psi_{ae}=-\bigl({3 \over \Lambda}\delta_{ae}+\epsilon_{ae}\bigr),
\end{equation}

\noindent
where $\epsilon_{ae}$ is the CDJ deviation matrix $\epsilon_{ae}$, which parametrizes deviations from $\boldsymbol{\psi}_{Kod}$.  Substitution of (\ref{KODD}) into (\ref{THEACTION5}) yields a canonical one form 

\begin{eqnarray}
\label{ONEFOR2}
\boldsymbol{\theta}_{Inst}=-{3 \over \Lambda}\int{dt}\int_{\Sigma}d^3xB^i_a\dot{A}^a_i
-\int{dt}\int_{\Sigma}d^3x\epsilon_{ae}B^i_e\dot{A}^a_i.
\end{eqnarray}

\noindent
The first term of (\ref{ONEFOR2}) is the integral of a total derivative, which integrates to the Chern--Simons functional $I_{CS}$.\par
\indent
When one expands the constraints relative to $\boldsymbol{\psi}_{Kod}$, one finds that the $-{3 \over \Lambda}\delta_{ae}$ part of (\ref{KODD}) drops out for the constraints linear in $\Psi_{ae}$.  So for the Gauss' law constraint we have

\begin{eqnarray}
\label{KODD3}
B^i_eD_i\Psi_{ae}=-B^i_eD_i\bigl({3 \over \Lambda}\delta_{ae}+\epsilon_{ae}\bigr)=-B^i_eD_i\epsilon_{ae},
\end{eqnarray}

\noindent
since the homogeneous and isotropic part of the CDJ matrix is annihilated by the covariant derivative.  Likewise, for the diffeomorphism constraint the $-{3 \over \Lambda}\delta_{ae}$ part cancels out due to antisymmetry

\begin{eqnarray}
\label{KODD5}
H_i=-\epsilon_{ijk}B^j_aB^k_e\bigl({3 \over \Lambda}\delta_{ae}+\epsilon_{ae}\bigr)=\epsilon_{ijk}B^j_aB^k_e\epsilon_{ae}=0.
\end{eqnarray}

\noindent
For the Hamiltonian constraint, an imprint of $-{3 \over \Lambda}\delta_{ae}$ remains upon expansion due to the nonlinearity of the contraint.  This can be seen as the imprint of $\boldsymbol{\psi}_{Kod}$, which interacts with the fluctuations.  The Hamiltonian constraint uses the invariants of the CDJ matrix, namely the trace 

\begin{eqnarray}
\label{THETRACE}
\hbox{tr}\Psi=-\bigl({9 \over \Lambda}+\hbox{tr}\epsilon\bigr)
\end{eqnarray}

\noindent
and the variance $Var\Psi$, given by

\begin{eqnarray}
\label{VAR}
Var\Psi=\epsilon_{abc}\epsilon_{efc}\bigl({3 \over \Lambda}\delta_{ae}+\epsilon_{ae}\bigr)\bigl({3 \over \Lambda}\delta_{bf}+\epsilon_{bf}\bigr)\nonumber\\
=\epsilon_{abc}\epsilon_{efc}\Bigl({9 \over {\Lambda^2}}\delta_{ae}\delta_{bf}
+{6 \over {\Lambda}}\delta_{ae}\epsilon_{bf}+\epsilon_{ae}\epsilon_{bf}\Bigr)
={{54} \over {\Lambda^2}}+{{12} \over {\Lambda}}\hbox{tr}\epsilon+Var\epsilon
\end{eqnarray}
 
\noindent
and the deteminant given by

\begin{eqnarray}
\label{DET} 
-6\hbox{det}\Psi=\epsilon_{abc}\epsilon_{efg}\bigl({3 \over \Lambda}\delta_{ae}+\epsilon_{ae}\bigr)
\bigl({3 \over \Lambda}\delta_{bf}+\epsilon_{bf}\bigr)\bigl({3 \over \Lambda}\delta_{cg}+\epsilon_{cg}\bigr)\nonumber\\
=\epsilon_{abc}\epsilon_{efg}\Bigl({{54} \over {\Lambda^3}}\delta_{ae}\delta_{bf}\delta_{cg}
+{{27} \over {\Lambda^2}}\delta_{ae}\delta_{bf}\epsilon_{cg}
+{9 \over {\Lambda}}\delta_{ae}\epsilon_{bf}\epsilon_{cg}+\epsilon_{ae}\epsilon_{bf}\epsilon_{cg}\Bigr)\nonumber\\
={{162} \over {\Lambda^3}}+{{54} \over {\Lambda^2}}\hbox{tr}\epsilon+{9 \over {\Lambda}}Var\epsilon+6\hbox{det}\epsilon
\end{eqnarray}

\noindent
Combining (\ref{VAR}) and (\ref{DET}), then the Hamiltonian constraint is given by

\begin{eqnarray}
\label{HAMI}
\hbox{det}B\bigl(\Lambda\hbox{det}\Psi+{1 \over 2}Var\Psi\bigr)\nonumber\\
=-\hbox{det}B\Bigl({6 \over {\Lambda}}\hbox{tr}\epsilon+2Var\epsilon
+2\Lambda\hbox{det}\epsilon\Bigr)=0.
\end{eqnarray}

\noindent
\indent
At the classical level, the constraints can be written as a system of seven equations in nine unknowns

\begin{eqnarray}
\label{CONSTRAINTTT}
\epsilon_{ijk}B^j_aB^k_e\epsilon_{ae}=0;~~
\textbf{w}_e\{\epsilon_{ae}\}=0;~~
\hbox{tr}\epsilon+{\Lambda \over 3}Var\epsilon+{{\Lambda^2} \over 3}\hbox{det}\epsilon=0.
\end{eqnarray}

\noindent
The third equation of (\ref{CONSTRAINTTT}) has used $(\hbox{det}B)\neq{0}$, which is a required condition for the transformation (\ref{THEACTION4}) to be valid.  Therefore the analysis of this paper does not apply to flat spacetimes, where $B^i_a=0$.

\section{Part One: The linearized theory}

Having expanded $\Psi_{ae}$ as in (\ref{KODD}), we will now linearize the theory using the following expansion about a reference connection $\alpha^a_i$ 

\begin{eqnarray}
\label{NOWEXPAND}
A^a_i=\alpha^a_i+a^a_i,
\end{eqnarray}

\noindent
where $\vert{a}^a_i\vert<<\alpha^a_i$.  We must substitute (\ref{NOWEXPAND}) into (\ref{ONEFOR2}) and (\ref{CONSTRAINTTT}) and expand to linear order in $a^a_i$.  The Ashtekar $SO(3,C)$ magnetic 
field $B^i_a=\epsilon^{ijk}\partial_jA^a_k+{1 \over }2\epsilon^{ijk}f_{abc}A^b_jA^c_k$ is given by

\begin{eqnarray}
\label{NOWEXPAND1}
B^i_a=\epsilon^{ijk}\partial_j(\alpha^a_k+a^a_k)+{1 \over 2}\epsilon^{ijk}f^{abc}(\alpha^b_j+a^b_j)(\alpha^c_k+a^c_k)\nonumber\\
=\beta^i_a[\alpha]+\epsilon^{ijk}(\partial_ja^a_k+f^{abc}\alpha^b_ja^c_k)+O(a^2).
\end{eqnarray}

\noindent
To make the physical content of the theory clear we will choose a reference connection $\alpha^a_i=\delta^a_i\alpha$, where $\alpha$ is a numerical constant.  Then we have

\begin{eqnarray}
\label{NOWEXPAND2}
B^i_a=\delta^i_a\alpha^2+\epsilon^{ijk}\partial_ja^a_k+\alpha(\delta^{ia}a^c_c-a^i_a)+\dots;~~C_{ae}=\delta_{ae}\alpha^3+\dots,
\end{eqnarray}

\noindent
where the dots signify higher order terms.  The canonical one form to linearized level, the second term of (\ref{ONEFOR2}), is given by

\begin{eqnarray}
\label{NOWEXPAND3}
\boldsymbol{\theta}_{Linear}=-{i \over G}\int_{\Sigma}d^3x\epsilon_{ae}B^i_e\dot{A}^a_i\nonumber\\
=-{i \over G}\int_{\Sigma}d^3x\epsilon_{ae}(\delta^i_e\alpha^2+\dots)(\dot{a}^a_i+\dots)
=-{i \over G}\alpha^2\int_{\Sigma}d^3x\epsilon_{ae}\dot{a}^a_e.
\end{eqnarray}

\noindent
To linearized order the theory exhibits a symplectic two form

\begin{eqnarray}
\label{NOWEXPAND31}
\boldsymbol{\Omega}_{Linear}=-{i \over G}\alpha^2\int_{\Sigma}d^3x{\delta\epsilon_{ae}}\wedge{\delta{a}_{ae}}\nonumber\\
=-{i \over G}\alpha^2\delta\Bigl(\int_{\Sigma}d^3x\epsilon_{ae}\delta{a}_{ae}\Bigr)=\delta\boldsymbol{\theta}_{Linear}.
\end{eqnarray}

\noindent
So at the unconstrained level one can read off the following elementary Poisson brackets from (\ref{NOWEXPAND31})

\begin{eqnarray}
\label{NOWEXPAND32}
\{a_{ae}(x,t),\epsilon_{bf}(y,t)\}=-i\Bigl({{\alpha^2} \over G}\Bigr)\delta_{ab}\delta_{ef}\delta^{(3)}(x,y).
\end{eqnarray}

\noindent
Since the constraints (\ref{CONSTRAINTTT}) are already of at least linear order in $\epsilon_{ae}$, then we need only expand them to zeroth order in $B^i_a$.  Hence the diffeomorphism constraint is given by

\begin{eqnarray}
\label{NOWEXPAND4}
H_i=\epsilon_{ijk}(\alpha^4\delta^j_a\delta^k_e)\epsilon_{ae}=\alpha^4\epsilon_{iae}\epsilon_{ae}=0,
\end{eqnarray}

\noindent
which implies that $\epsilon_{ae}=\epsilon_{ea}$ must be symmetric.  The Hamiltonian constraint to linearized order is given by

\begin{eqnarray}
\label{NOWEXPAND5}
\hbox{tr}\epsilon=0,
\end{eqnarray}

\noindent
which states that $\epsilon_{ae}$ is traceless to this order.  For the Gauss' law constraint we have

\begin{eqnarray}
\label{NOWEXPAND6}
G_a=\alpha^2\delta^i_e\partial_i\epsilon_{ae}+\alpha^3\delta_{be}\bigl(f_{abf}\delta_{ge}+f_{ebg}\delta_{af}\bigr)\epsilon_{fg}\nonumber\\
=\alpha^2\partial_e\epsilon_{ae}+\alpha^3f_{agf}\epsilon_{fg}=0.
\end{eqnarray}

\noindent
The second term on the right hand side of (\ref{NOWEXPAND6}) vanishes since $\epsilon_{ae}$ is symmetric from (\ref{NOWEXPAND4}), and the Gauss' law constraint reduces to

\begin{eqnarray}
\label{NOWEXPAND7}
\partial_e\epsilon_{ae}=0,
\end{eqnarray}

\noindent
which states that $\epsilon_{ae}$ is transverse.  Since upon implementation of the linearized constraints $\epsilon_{ae}$ is symmetric, traceless and transverse then it corresponds to a spin two field.

\subsection{Massless spin two polarizations}

We will now make contact with the conventional formalism, as is best seen in momentum space, using a plane waveform for $\epsilon_{ae}$.  From (\ref{NOWEXPAND4}) and (\ref{NOWEXPAND5}) the most general form for $\epsilon_{ae}$ is given by the parametrization of its diagonal and off-diagonal parts, $\varphi_f$ and $\Psi_f$ respectively\footnote{We have omitted the time dependence, since the initial value constraints are solved with respect to a given spatial hypersurface $\Sigma_t$ for each time $t$.}

\begin{displaymath}
\epsilon_{ae}=
\left(\begin{array}{ccc}
\varphi_1 & \Psi_3 & \Psi_2\\
\Psi_3 & \varphi_2 & \Psi_1\\
\Psi_2 & \Psi_1 & \varphi_3\\
\end{array}\right)
e^{\vec{k}\cdot\vec{r}},
\end{displaymath}
 
\noindent
subject to the tracelessness condition $\hbox{tr}\epsilon=\varphi_1+\varphi_2+\varphi_3=0$, where $\vec{k}=(k_1,k_2,k_3)$ is the wave vector of the gravitational wave.  The linearized Gauss' law constraint (\ref{NOWEXPAND7}) is given by

\begin{displaymath}
k_e\epsilon_{ae}=
\left(\begin{array}{ccc}
\varphi_1 & \Psi_3 & \Psi_2\\
\Psi_3 & \varphi_2 & \Psi_1\\
\Psi_2 & \Psi_1 & \varphi_3\\
\end{array}\right)
\left(\begin{array}{c}
k_1\\
k_2\\
k_3\\
\end{array}\right)
=
\left(\begin{array}{c}
0\\
0\\
0\\
\end{array}\right)
,
\end{displaymath}

\noindent
which can be rewritten as

\begin{displaymath}
\left(\begin{array}{ccc}
0 & k_3 & k_2\\
k_3 & 0 & k_1\\
k_2 & k_1 & 0\\
\end{array}\right)
\left(\begin{array}{c}
\Psi_1\\
\Psi_2\\
\Psi_3\\
\end{array}\right)
=-
\left(\begin{array}{ccc}
k_1 & 0 & 0\\
0 & k_2 & 0\\
0 & 0 & k_3\\
\end{array}\right)
\left(\begin{array}{c}
\varphi_1\\
\varphi_2\\
\varphi_3\\
\end{array}\right)
.
\end{displaymath}

\noindent
To make the physical content more apparent in terms of gravitation modes, let us use a wave vector of the form $\vec{k}=(k_1,0,0)$, which corresponds to a wave travelling 
in the $\textbf{x}$ direction of a Cartesian coordinate system.  For $k_2=k_3=0$ this is given in matrix form by

\begin{displaymath}
\left(\begin{array}{ccc}
0 & 0 & 0\\
0 & 0 & k_1\\
0 & k_1 & 0\\
\end{array}\right)
\left(\begin{array}{c}
\Psi_1\\
\Psi_2\\
\Psi_3\\
\end{array}\right)
=-
\left(\begin{array}{ccc}
k_1 & 0 & 0\\
0 & 0 & 0\\
0 & 0 & 0\\
\end{array}\right)
\left(\begin{array}{c}
\varphi_1\\
\varphi_2\\
\varphi_3\\
\end{array}\right)
,
\end{displaymath}

\noindent
This yields the equations

\begin{eqnarray}
\label{WAVE30}
0=\varphi_1k_1;~~k_3\Psi_3=0;~~k_1\Psi_2=0,
\end{eqnarray}

\noindent
from which we have that $\varphi_1=\Psi_2=\Psi_3=0$.  But since $\epsilon_{ae}$ is traceless with $\varphi_1=0$, then $\varphi_3=-\varphi_2$.  The deviation matrix is then of the form

\begin{displaymath}
\epsilon_{ae}=\varphi
\left(\begin{array}{ccc}
0 & 0 & 0\\
0 & 1 & 0\\
0 & 0 & -1\\
\end{array}\right)
e^{\vec{k}\cdot\vec{r}}+\Psi
\left(\begin{array}{ccc}
0 & 0 & 0\\
0 & 0 & 1\\
0 & 1 & 0\\
\end{array}\right)
e^{\vec{k}\cdot\vec{r}}.
\end{displaymath}

\noindent
We have obtained the two polarizations of a massless spin two field in $SO(3,C)$ language.  A similar result can be obtained for waves travelling in the  $\textbf{y}$ and the $\textbf{z}$ directions.\par
\indent
For the general case where the wave vector $\vec{k}$ is not aligned with the coordinate directions the, Gauss' law constraint can be written as

\begin{eqnarray}
\label{WAVE28}
\Psi_f=\hat{J}^g_f\varphi_g.
\end{eqnarray}

\noindent
This expresses the off-diagonal elements $\Psi_f$ as the image of the diagonal elements $\varphi_f$ with respect to a propagator $\hat{J}^g_f$.  When $\hat{J}^g_f$ exists, then equation (\ref{WAVE28}) in matrix form is given by

\begin{displaymath}
\left(\begin{array}{c}
\Psi_1\\
\Psi_2\\
\Psi_3\\
\end{array}\right)
=-
\left(\begin{array}{ccc}
0 & k_3 & k_2\\
k_3 & 0 & k_1\\
k_2 & k_1 & 0\\
\end{array}\right)^{-1}
\left(\begin{array}{ccc}
k_1 & 0 & 0\\
0 & k_2 & 0\\
0 & 0 & k_3\\
\end{array}\right)
\left(\begin{array}{c}
\varphi_1\\
\varphi_2\\
\varphi_3\\
\end{array}\right)
,
\end{displaymath}

\noindent
which hinges upon the ability to invert the off-diagonal matrix of wave vector components.  This is given by

\begin{displaymath}
\left(\begin{array}{c}
\Psi_1\\
\Psi_2\\
\Psi_3\\
\end{array}\right)
=(2k_1k_2k_3)^{-1}
\left(\begin{array}{ccc}
-k_1^3 & k_1k_2^2 & k_3^2k_1\\
k_1^2k_2 & -k_2^3 & k_2k_3^2\\
k_3^2k_1 & k_2^2k_3 & -k_3^3\\
\end{array}\right)
\left(\begin{array}{c}
\varphi_1\\
\varphi_2\\
\varphi_3\\
\end{array}\right)
,
\end{displaymath}

\noindent
whence one sees that we must have $k_1k_2k_3\neq{0}$.  Note that $\hat{J}^g_f$ does not exist for the previous case of propagation along the coordinate directions.\par
\indent
This solution for $\epsilon_{ae}$ contains two degrees of freedom per point and can be written completely in terms of the traceless diagonal elements $\varphi_g$ via the relation

\begin{eqnarray}
\label{WAVE29}
\epsilon_{ae}=\bigl((e^g)_{ae}+(E^f)_{ae}\hat{J}^g_f\bigr)\varphi_g\equiv(\hat{T}^g)_{ae}\varphi_g.
\end{eqnarray}

\noindent
In equation (\ref{WAVE29}), $\hat{T}^g_{ae}$ is an operator which implements an embedding map from the two dimensional space $(\varphi_1,\varphi_2)$ into the six dimensional space $\epsilon_{ae}$, taking the kinematics of the Gauss' law constraint into account.

\subsection{Quantization and Hilbert space of the linearized theory}

\noindent
We will now perform a quantization by promoting Poisson brackets (\ref{NOWEXPAND32}) to equal-time commutators

\begin{eqnarray}
\label{NOWEXPAND32}
\bigl[\hat{a}_{ae}(x,t),\hat{\epsilon}_{bf}(y,t)\bigr]=\alpha^{-2}(\hbar{G})\delta_{ab}\delta_{ef}\delta^{(3)}(x,y).
\end{eqnarray}

\noindent
Since the variables are complex, then to ensure square integrability of the wavefunctions of the auxilliary Hilbert space we will use a Gaussian measure

\begin{eqnarray}
\label{NOWEXPAND321}
D\mu=\prod_{x,a,e}\delta{a}_{ae}(x)\hbox{exp}\Bigl[-\mu\int_{\sigma}d^3x\overline{a}_{ae}(x)a_{ae}(x)\Bigr]
\end{eqnarray}

\noindent
for normalization, where $\mu$ is a numerical constant of mass dimension $[\mu]=1$.  For the auxilliary Hilbert space we will use holomorphic plane waves $\boldsymbol{\psi}$ which are eigenstates of the momentum operators, given by

\begin{eqnarray}
\label{NOWEXPAND322}
\boldsymbol{\psi}_{\lambda}[a]=\hbox{exp}\Bigl[-\alpha^4\mu^{-1}(\hbar{G})^{-2}\int_{\Sigma}d^3x\lambda^{*}_{ae}(x)\lambda_{ae}(x)\Bigr]\hbox{exp}\Bigl[\alpha^2(\hbar{G})^{-1}\int_{\Sigma}d^3x\lambda_{ae}(x)a_{ae}(x,t)\Bigr],
\end{eqnarray}

\noindent
where the pre-factor is a normalization factor and $\lambda_{ae}$ labels the state.  The action of the operators on (\ref{NOWEXPAND321}) are given by

\begin{eqnarray}
\label{NOWEXPAND323}
\hat{a}_{ae}(x,t)\boldsymbol{\psi}_{\lambda}=a_{ae}(x,t)\boldsymbol{\psi}_{\lambda};~~
\hat{\epsilon}_{ae}(x,t)\boldsymbol{\psi}=\alpha^{-2}(\hbar{G}){\delta \over {\delta{a}_{ae}(x,t)}}\boldsymbol{\psi}_{\lambda}.
\end{eqnarray}

\noindent
There are two possibilities for quantization of gravitons, depending on whether the Gauss' law propagator $\hat{J}^f_g$ exists or not.  In the latter case one may parametrize the momentum space degrees of freedom by

\begin{displaymath}
\epsilon_{ae}=\pi_1
\left(\begin{array}{ccc}
0 & 0 & 0\\
0 & 1 & 0\\
0 & 0 & -1\\
\end{array}\right)
+\pi_2
\left(\begin{array}{ccc}
0 & 0 & 0\\
0 & 0 & 1\\
0 & 1 & 0\\
\end{array}\right)
.
\end{displaymath}

\noindent
Since the constraints (\ref{CONSTRAINTTT}) do not constrain the configuration space then we are free to choose $a_{ae}$, each choice tantamount to the choice of a gauge.  Let us make choose the connection 

\begin{displaymath}
a_{ae}=a_1
\left(\begin{array}{ccc}
0 & 0 & 0\\
0 & 1 & 0\\
0 & 0 & -1\\
\end{array}\right)
+a_2
\left(\begin{array}{ccc}
0 & 0 & 0\\
0 & 0 & 1\\
0 & 1 & 0\\
\end{array}\right)
.
\end{displaymath}

\noindent
Then the commutation relations (\ref{NOWEXPAND2}) reduce to

\begin{eqnarray}
\label{NOWEXPAND324}
\bigl[\hat{a}_1(x,t),\hat{\pi}_1(y,t)\bigr]=\bigl[\hat{a}_2(x,t),\hat{\pi}_2(y,t)\bigr]=\alpha^{-2}(\hbar{G})\delta^{(3)}(x,y).
\end{eqnarray}

\noindent
But since the phase space must have six dimensions per point at the level prior to implementation of the Hamiltonian constraint, we need a third variable $a$ with conjugate momentum $\pi=\hbox{tr}\epsilon$ satisfying the relation

\begin{eqnarray}
\label{NOWEXPAND38}
\bigl[\hat{a}(x,t),\hat{\pi}(y,t)\bigr]=\alpha^{-2}(\hbar{G})\delta^{(3)}(x,y).
\end{eqnarray}

\noindent
This enables us to directly implement the Hamiltonian constraint in the more general case where $\hat{J}_f^g$ exists, the Hamiltonian constraint can be implemented at the quantum level completely in terms of the diagonal 
elements $\varphi_f$.  The quantum Hamiltonian constraint to linearized order is then given by

\begin{eqnarray}
\label{NOWEXPAND39}
\hat{H}\boldsymbol{\psi}=\alpha^{-2}(\hbar{G}){\delta \over {\delta{a}(x,t)}}\boldsymbol{\psi}=0,
\end{eqnarray}

\noindent
which states that $\boldsymbol{\psi}$ is independent of $a$.  If we interpret $a$ as a time variable on configuration space, then this means that $\boldsymbol{\psi}$ is independent of time.  Then the most general 
solution is $\boldsymbol{\psi}[a_1,a_2]$, which depends on the two physical degrees of freedom which are orthogonal to the time direction.  In this case the wavefunctionals solving the constraints 
are given by $\boldsymbol{\psi}[a_1,a_2]\in{L}^2(a_1,a_2;D\mu)$, the set of square integrable functions of $a_1$ and $a_2$ in the measure (\ref{NOWEXPAND321}).\par
\indent

\section{Part Two: the full nonlinear theory}

Having demonstrated the existence of a well-defined Hilbert space structure for complex gravity at the linearized level on the phase space $\Omega_{Inst}$, we will now demonstrate the same for the full, nonlinear theory.\footnote{The intent is to show that the physical degrees of freedom of the full theory are preserved under linearization.}  First, we must show that the full nonlinearized version of the constraints (\ref{CONSTRAINTTT}) admit a solution at the classical level.  The diffeomorphism constraint is

\begin{eqnarray}
\label{NONLIN}
H_i=\epsilon_{ijk}B^j_kB^a_e\Psi_{ae}=(\hbox{det}B)(B^{-1})^f_i\epsilon_{fae}\Psi_{ae}=0.
\end{eqnarray}

\noindent
Since $(\hbox{det}B)\neq{0}$ by assumption, then equation (\ref{NONLIN}) states that $\Psi_{ae}=\Psi_{ea}$ is symmetric.  This is the case independently of linearization and holds for all 
connections $A^a_i$ with $(\hbox{det}B)\neq{0}$.  According to \cite{WEYL}, a complex symmetric 3 by 3 matrix can be diagonalized when there exist three linearly independent eigenvectors.  In the case of $\epsilon_{ae}$ this is the case when $(\hbox{det}\epsilon)\neq{0}$, which enables us to write the following polar decomposition

\begin{eqnarray}
\label{NONLIN1}
\epsilon_{ae}=(e^{\theta\cdot{T}})_{af}\lambda_f(e^{-\theta\cdot{T}})_{fe}.
\end{eqnarray}

\noindent
In equation (\ref{NONLIN1}) $\vec{\theta}=(\theta^1,\theta^2,\theta^3)$ are three complex rotation parameters, which implement a transformation of the eigenvalues $\lambda_f=(\lambda_1,\lambda_2,\lambda_3)$ into a new Lorentz frame.\par
\indent
On account of the cyclic property of the trace, the Hamiltonian constraint can now be written completely in terms of the eigenvalues 

\begin{eqnarray}
\label{NONLIN2}
H=\hbox{tr}\epsilon+{\Lambda \over 3}Var\epsilon+{{\Lambda^2} \over 3}\hbox{det}\epsilon=\lambda_1+\lambda_2+\lambda_3\nonumber\\
+{{2\Lambda} \over 3}\bigl(\lambda_1\lambda_2+\lambda_2\lambda_3+\lambda_3\lambda_1\bigr)+{{\Lambda^2} \over 3}\lambda_1\lambda_2\lambda_3=0.
\end{eqnarray}

\noindent
Just as we applied the quantization procedure to $\hbox{tr}\epsilon=0$ in the linerized theory, we will as well apply the quantization procedure to (\ref{NONLIN2}), which brings into question the role of the Gauss' law constraint.  Using the parametrization (\ref{NONLIN1}) the full nonlinear Gauss' law constraint can be written as

\begin{eqnarray}
\label{NONLIN3}
G_a=\textbf{w}_e\{\lambda_f(e^{-\theta\cdot{T}})_{fa}(e^{-\theta\cdot{T}})_{fe}\}=0.
\end{eqnarray}

\noindent
Note that (\ref{NONLIN2}) is independent of $\vec{\theta}$ and depends only on $\lambda_f$, which we will regard as the physical degrees of freedom.  From this perspective, we will regard (\ref{NONLIN3}) as a condition for determining the angles $\vec{\theta}$.  The set of connections $A^a_i$ defines an equivalence class of angles $\vec{\theta}\equiv\vec{\theta}[\vec{\lambda};A]$ labelled by each 
triple of eigenvalues $\lambda_f$ satisfying (\ref{NONLIN2}).\par
\indent
All that remains then is to show that the canonical structure upon the indentification (\ref{NONLIN1}) reduces accordingly to a canonical structure on the reduced phase space under $H_i$ and $G_a$.  This can be seen from the relation 

\begin{eqnarray}
\label{NONLIN4}
\epsilon_{ae}B^i_e\dot{A}^a_i=\lambda_f((e^{-\theta\cdot{T}})_{fe}B^i_e)((e^{-\theta\cdot{T}})_{fa}\dot{A}^a_i,
\end{eqnarray}

\noindent
whence the $SO(3,C)$ matrices rotate the internal indices of $B^i_a$ and $\dot{A}^a_i$.  Since the velocity $\dot{A}^a_i$ lives in the tangent space to configuration space, then it transforms the same way as $B^i_a$ under $SO(3,C)$ gauge transformations, namely inhomogeneously.  We can then make the identifications

\begin{eqnarray}
\label{NONLIN5}
\dot{a}^a_i=(e^{-\theta\cdot{T}})_{fa}\dot{A}^a_i;~~b^i_a=(e^{-\theta\cdot{T}})_{fe}B^i_a,
\end{eqnarray}

\noindent
and regard the new connection $a^a_i$ as specially adapted to an `intrinsic' $SO(3,C)$ frame corresponding to the eigenvalues $\lambda_f$.  So we can redefine a new theory, starting at the level after implementation of the diffeomorphism and Gauss' law constraints, with canonical one form

\begin{eqnarray}
\label{NONLIN6}
\boldsymbol{\theta}_{Kin}=\int_{\Sigma}d^3x\lambda_fb^i_f\delta{a}^f_i
\end{eqnarray}

\noindent
where the angles $\vec{\theta}$ are ignorable.  However, (\ref{NONLIN6}) is presently not in a form suitable for quantization.  This is because its functional variation

\begin{eqnarray}
\label{THEACTION9}
\delta\boldsymbol{\theta}_{Kin}=
=\int_{\Sigma}d^3x\Bigl[b^i_f{\delta\lambda_f}\wedge{\delta{a}^f_i}+\lambda_f\epsilon^{ijk}{(D_j\delta{a}^f_k)}\wedge{\delta{a}^f_i}\Bigr],
\end{eqnarray}

\noindent
where $D_i\equiv(D^{ae})_i=\delta^{ae}\partial_i+f^{abe}a^b_i$ is the covariant derivative with respect to the connection $a^f_i$, in direct analogy to (\ref{THEACTION9}) is not of symplectic form owing to the presence of the second term.  Additionally, there is a mismatch in degrees of freedom between the momentum space and the configurations space.  To have a cotangent bundle structure, we need three configuration space degrees of freedom corresponding to the eigenvalues $\lambda_f$.  Since the constraints do not place any restriction on the connection $A^a_i$, then provided $\hbox{det}\neq{0}$ we are free the choose any connection we wish.  For the purposes of this paper we will 
limit ourself to a diagonal connection $A^a_i=\delta^a_iA^a_a$ with no summation over $a$.  We will see that this choice eliminates the second term of (\ref{THEACTION9}), while providing the requisite canonical structure necessary for quantization of the full, nonlinear theory.

\subsection{Canonical structure}

\noindent
We will now put in place the canonical structure required to quantize the fluctuations about the Kodama state.  After rotation of all variables into the intrinsic $SO(3,C)$ frame, one sees that a choice of diagonal configuration space variables admits a canonical structure.  The canonical one form is given by\footnote{We have identified the diagonal elements of $\epsilon_{ae}$ with its eigenvalues, which means that the variables have been adapted 
to a $SO(3,C)$ frame solving the Gauss' law constraint.}

\begin{eqnarray}
\label{JACOBI}
\boldsymbol{\theta}_{Kin}={i \over G}\int_{\Sigma}d^3x\epsilon_{ae}B^i_e\delta{A}^a_i\nonumber\\
={i \over G}\int_{\Sigma}d^3x\biggl(\epsilon_{11}A^2_2A^3_3\delta{A}^1_1+\epsilon_{22}A^3_3A^1_1\delta{A}^2_2+\epsilon_{33}A^1_1A^2_2\delta{A}^3_3\biggr).
\end{eqnarray}

\noindent
Now define densitized momentum variables $\widetilde{\epsilon}_{ae}=\epsilon_{ae}(\hbox{det}A)$, where $(\hbox{det}A)\neq{0}$.  Hence we have
 
\begin{eqnarray}
\label{JACOBI1}
\widetilde{\epsilon}_{11}=\epsilon_1(A^1_1A^2_2A^3_3);~~\widetilde{\epsilon}_{22}=\epsilon_2(A^1_1A^2_2A^3_3);~~\widetilde{\epsilon}_{33}=\epsilon_3(A^1_1A^2_2A^3_3).
\end{eqnarray}

\noindent
In the densitized variables (\ref{JACOBI1}), then (\ref{JACOBI}) is given by 

\begin{eqnarray}
\label{JACOBI2}
\boldsymbol{\theta}_{Kin}={i \over G}\int_{\Sigma}d^3x
\biggl(\widetilde{\epsilon}_{11}\Bigl({{\delta{A}^1_1} \over {A^1_1}}\Bigr)+
\widetilde{\epsilon}_{22}\Bigl({{\delta{A}^2_2} \over {A^2_2}}\Bigr)+
\widetilde{\epsilon}_{33}\Bigl({{\delta{A}^3_3} \over {A^3_3}}\Bigr)\biggr).
\end{eqnarray}

\noindent
Next, rewrite (\ref{JACOBI2}) in the form

\begin{eqnarray}
\label{JACOBI3}
\boldsymbol{\theta}_{Kin}={i \over G}\int_{\Sigma}d^3x
\biggl((\widetilde{\epsilon}_{11}-\widetilde{\epsilon}_{33}){{\delta{A}^1_1} \over {A^1_1}}+
(\widetilde{\epsilon}_{22}-\widetilde{\epsilon}_{33}){{\delta{A}^2_2} \over {A^2_2}}\nonumber\\
+\widetilde{\epsilon}_{33}\Bigl({{\delta{A}^1_1} \over {A^1_1}}+{{\delta{A}^2_2} \over {A^2_2}}+{{\delta{A}^3_3} \over {A^3_3}}\Bigr)\biggr)
\end{eqnarray}

\noindent
and make the following change of variables

\begin{eqnarray}
\label{JACOBI4}
\Pi_1=\Bigl({\Lambda \over {3a_0^3}}\Bigr)(\widetilde{\epsilon}_{11}-\widetilde{\epsilon}_{33});~~
\Pi_2=\Bigl({\Lambda \over {3a_0^3}}\Bigr)(\widetilde{\epsilon}_{22}-\widetilde{\epsilon}_{33});~~\Pi=\Pi_1=\Bigl({\Lambda \over {3a_0^3}}\Bigr)\widetilde{\epsilon}_{33}
\end{eqnarray}

\noindent
where $a_0$ is a numerical constant of mass dimension $[a_0]=1$.  Then for the configuration space make the definition

\begin{eqnarray}
\label{JACOBI5}
{{\delta{A}^1_1} \over {A^1_1}}=\delta{X};~~{{\delta{A}^2_2} \over {A^2_2}}=\delta{Y};~~{{\delta{A}^1_1} \over {A^1_1}}+{{\delta{A}^2_2} \over {A^2_2}}+{{\delta{A}^3_3} \over {A^3_3}}=\delta{T}.
\end{eqnarray}

\noindent
Equation (\ref{JACOBI5}) provides global coordinates $(X,Y,T)$ on the kinematic onfiguration space $\Gamma_{Kin}$, given by

\begin{eqnarray}
\label{JACOBI6}
X=\hbox{ln}\Bigl({{A^1_1} \over {a_0}}\Bigr);~~Y=\hbox{ln}\Bigl({{A^2_2} \over {a_0}}\Bigr);~~T=\hbox{ln}\Bigl({{A^1_1A^2_2A^3_3} \over {a_0^3}}\Bigr).
\end{eqnarray}

\noindent
The canonical one form corresponding to (\ref{JACOBI7}) is given by  

\begin{eqnarray}
\label{ONEFORM7}
\boldsymbol{\theta}_{Kin}=\Bigl({{3ia_0^3} \over {G\Lambda}}\Bigr)\int_{\Sigma}d^3x\int{dt}\bigl(\Pi\dot{T}+\Pi_1\dot{X}+\Pi_2\dot{Y}\bigr).
\end{eqnarray}

\noindent
With the variables as defined, (\ref{ONEFORM7}) yields a symplectic two form

\begin{eqnarray}
\label{JACOBI7}
\boldsymbol{\Omega}_{Kin}
=\Bigl({{3ia_0^3} \over {G\Lambda}}\Bigr)\int_{\Sigma}d^3x\Bigl({\delta\Pi_1}\wedge{\delta{X}}+{\delta\Pi_2}\wedge{\delta{Y}}+{\delta\Pi}\wedge{\delta{T}}\Bigr)=\delta\boldsymbol{\theta}_{Kin}.
\end{eqnarray}

\noindent
The mass dimensions of the dynamical variables are 

\begin{eqnarray}
\label{JACOBI8}
[\Pi_1]=[\Pi_2]=[\Pi]=[X]=[Y]=[T]=0, 
\end{eqnarray}

\noindent
and we have the Poisson brackets

\begin{eqnarray}
\label{BRACK}
\{\hat{T}(x,t),\hat{\Pi}(y,t)\}=\{\hat{X}(x,t),\hat{\Pi}_1(y,t)\}=\{\hat{Y}(x,t),\hat{\Pi}_2(y,t)\}=\Bigl({{G\Lambda} \over {3ia_0^3}}\Bigr)\delta^{(3)}(x,y)
\end{eqnarray}

\noindent
The advantage of the choice of dimensionless variables is that the Hamiltonian constraint, the third equation of (\ref{CONSTRAINTTT}), can be written as a dimensionless equation.  First rewrite it in terms of the densitized variables

\begin{eqnarray}
\label{CANNOW}
(\hbox{det}A)^{-1}\hbox{tr}\widetilde{\epsilon}+{\Lambda \over 3}(\hbox{det}A)^{-2}Var\widetilde{\epsilon}+{{\Lambda^2} \over 3}(\hbox{det}A)^{-3}\hbox{det}\widetilde{\epsilon}=0.
\end{eqnarray}

\noindent
Multiplication of (\ref{CANNOW}) by ${\Lambda \over 3}(\hbox{det}A)^3a_0^{-9}$ and using (\ref{JACOBI4}) enables the Hamiltonian constraint to be written as

\begin{eqnarray}
\label{CANNOW1}
H=e^{2T}\bigl(3\Pi+\Pi_1+\Pi_2\bigr)+e^T\bigl(3\Pi^2+2(\Pi_1+\Pi_2)\Pi\nonumber\\
+\Pi_1\Pi_2\bigr)+3\Pi(\Pi+\Pi_1)(\Pi+\Pi_2)=0.
\end{eqnarray}

\noindent
Let us define the following operators

\begin{eqnarray}
\label{OPERATE}
Q^{(1)}=\Pi+{1 \over 3}(\Pi_1+\Pi_2);\nonumber\\
Q^{(2)}=\Pi^2+{2 \over 3}(\Pi_1+\Pi_2)\Pi+{1 \over 3}\Pi_1\Pi_2;\nonumber\\
O=\Pi(\Pi+\Pi_1)(\Pi+\Pi_2).
\end{eqnarray}

\noindent
Then upon dividing by a factor of $3$ the Hamiltonian constraint can be written as 

\begin{eqnarray}
\label{OPERATE1}
H=e^TQ^{(1)}+e^{2T}Q^{(2)}+O=0.
\end{eqnarray}

\subsection{Quantization and auxilliary Hilbert space}

To pass over into the quantum theory we promote Poisson brackets (\ref{BRACK}) to commutators.  Upon quantization, the variables $\Pi$, $\Pi_1$ and $\Pi_2$ become 
promoted to operators $\hat{\Pi}$, $\hat{\Pi}_1$, and $\hat{\Pi}_2$ and $T$, $X$ and $Y$ to operators $\hat{T}$, $\hat{X}$ and $\hat{Y}$ satisfying the nontrivial equal time commutation relations

\begin{eqnarray}
\label{QUANTIZATION}
\bigl[\hat{T}(x,t),\hat{\Pi}(y,t)\bigr]=\bigl[\hat{X}(x,t),\hat{\Pi}_1(y,t)\bigr]=\bigl[\hat{Y}(x,t),\hat{\Pi}_2(y,t)\bigr]=\mu\delta^{(3)}(x,y)
\end{eqnarray}

\noindent
where we have defined the constant

\begin{eqnarray}
\label{HAVEDEFINED}
\mu=\Bigl({{\hbar{G}\Lambda} \over {3a_0^3}}\Bigr).
\end{eqnarray}

\noindent
In the functional Schr\"odinger representation, holomorphic in $X$, $Y$ and $T$, the operators act respectively by multiplication 

\begin{eqnarray}
\label{QUANTUM1}
\hat{T}(x,t)\boldsymbol{\psi}=T(x,t)\boldsymbol{\psi};~~
\hat{X}(x,t)\boldsymbol{\psi}=X(x,t)\boldsymbol{\psi};~~
\hat{Y}(x,t)\boldsymbol{\psi}=Y(x,t)\boldsymbol{\psi}
\end{eqnarray}

\noindent
and by functional differentiation

\begin{eqnarray}
\label{QUANTUM2}
\hat{\Pi}(x,t)\boldsymbol{\psi}=\mu{\delta \over {\delta{T}(x,t)}}\boldsymbol{\psi};~~
\hat{\Pi}_1(x,t)\boldsymbol{\psi}=\mu{\delta \over {\delta{X}(x,t)}}\boldsymbol{\psi};~~
\hat{\Pi}_2(x,t)\boldsymbol{\psi}=\mu{\delta \over {\delta{Y}(x,t)}}\boldsymbol{\psi}.
\end{eqnarray}

\noindent
The wavefunction $\boldsymbol{\psi}$ is determined from the following resolution of the identity

\begin{eqnarray}
\label{QUANTUM3}
I=\prod_x\int{\delta\mu}\bigl\vert{X},Y,T\bigr>\bigl<X,Y,T\bigr\vert,
\end{eqnarray}

\noindent
whence the state diagonal in the configuration variables is given by

\begin{eqnarray}
\label{QUANTUM4}
\boldsymbol{\psi}(X,Y,T)=\bigl<X,Y,T\bigl\vert\boldsymbol{\psi}\bigr>.
\end{eqnarray}  

\noindent
Since the variables are complex, we choose a measure Gaussian in $X$ and $Y$ to ensure normalizable wavefunctions.  This is given by\footnote{Note, we do not include a measure in $T$, because we will interpret $T$ as a time variable and one does not normalize a wavefunction in time.}

\begin{eqnarray}
\label{NOMRAL}
D\mu(X,Y)=\prod_x\delta{X}\delta\overline{X}\delta{dY}\delta\overline{Y}\hbox{exp}\Bigl[-\nu^{-1}\int_{\Sigma}d^3x(\vert{X}\vert^2+\vert{Y}\vert^2)\Bigr]
\end{eqnarray}

\noindent
where $\nu$ is a numerical constant of mass dimension $[\nu]=-3$.  For the auxilliary Hilbert space we will use eigenstates of the momentum operators

\begin{eqnarray}
\label{EIGENSTATE}
\boldsymbol{\psi}=\bigl<X,Y,T\bigl\vert\alpha,\beta,\lambda\bigr>
=e^{\mu^{-1}(\alpha\cdot{X}+\beta\cdot{Y}+\lambda\cdot{T})},
\end{eqnarray}

\noindent
where the dot signifies an integration over 3-space $\Sigma$ as in

\begin{eqnarray}
\label{EIGENSTATE1}
U\cdot{V}=\int_{\Sigma}d^3xU(x)V(x).
\end{eqnarray}

\noindent
So we have that

\begin{eqnarray}
\label{EIGENSTATE2}
\hat{\Pi}_1\bigl\vert\alpha,\beta,\lambda\bigr>=\alpha\bigl\vert\alpha,\beta,\lambda\bigr>;~~
\hat{\Pi}_2\bigl\vert\alpha,\beta,\lambda\bigr>=\beta\bigl\vert\alpha,\beta,\lambda\bigr>;~~
\hat{\Pi}\bigl\vert\alpha,\beta,\lambda\bigr>=\lambda\bigl\vert\alpha,\beta,\lambda\bigr>.
\end{eqnarray}

\noindent
The overlap between two unnormalized states is given, using (\ref{NOMRAL}) for a measure, by

\begin{eqnarray}
\label{OVERLAP}
\bigl\vert\bigl<{\alpha,\beta}\bigl\vert{\alpha^{\prime},\beta^{\prime}}\bigr>\bigr\vert^2
=e^{-\nu\mu^{-2}\vert\alpha-\alpha^{\prime}\vert^2}e^{-\nu\mu^{-2}\vert\beta-\beta^{\prime}\vert^2}e^{\mu^{-1}(\lambda\cdot{T}+\lambda^{*}\cdot\overline{T})}.
\end{eqnarray}

\noindent
There is always a nontrivial overlap between the states, which is a consequence of Gaussian measure needed for the holomorphic representation.  Hence, the states as defined by (\ref{EIGENSTATE}) form an overcomplete set.\par
\indent
Note that the operators (\ref{OPERATE}) have the following action on the auxilliary states

\begin{eqnarray}
\label{EIGENSTATE3}
\hat{Q}^{(1)}\vert\alpha,\beta,\lambda\bigr>=\bigl(\lambda+{1 \over 3}(\alpha+\beta)\bigr)\bigl\vert\alpha,\beta,\lambda\bigr>;\nonumber\\
\hat{Q}^{(2)}\vert\alpha,\beta,\lambda\bigr>=(\lambda+\gamma^{-}(\alpha,\beta))(\lambda+\gamma^{+}(\alpha,\beta))\bigl\vert\alpha,\beta,\lambda\bigr>;\nonumber\\
\hat{O}\vert\alpha,\beta,\lambda\bigr>=\lambda(\lambda+\alpha)(\lambda+\beta)\bigl\vert\alpha,\beta,\lambda\bigr>,
\end{eqnarray}

\noindent
where $\gamma^{\pm}(\alpha,\beta)$ are the roots of

\begin{eqnarray}
\label{EIGENSTATE31}
\lambda^2+{2 \over 3}(\alpha+\beta)\lambda+{1 \over 3}\alpha\beta=0.
\end{eqnarray}

\noindent
For the quantum Hamiltonian constraint we will choose an operator ordering with the momenta to the left of the coordinates, as in

\begin{eqnarray}
\label{EIGENSTATE4}
\hat{H}=\hat{Q}^{(1)}e^T+\hat{Q}^{(2)}e^{2T}+\hat{O}.
\end{eqnarray}

\subsection{Construction of the states}

\noindent
The quantum Hamiltonian constraint can be written as

\begin{eqnarray}
\label{THEQUANTUM}
\hat{O}\bigl\vert\boldsymbol{\psi}\bigr>=-\bigl(\hat{Q}^{(1)}e^T+\hat{Q}^{(2)}e^{2T}\bigr)\bigl\vert\boldsymbol{\psi}\bigr>.
\end{eqnarray}

\noindent
Let there be states $\boldsymbol{\psi}_0\in{Ker}\{\hat{O}\}$.  Then acting on (\ref{THEQUANTUM}) with $\hat{O}^{-1}$, assumed to be invertible, we obtain

\begin{eqnarray}
\label{THEQUANTUM1}
\bigl\vert\boldsymbol{\psi}\bigr>=\bigl\vert\boldsymbol{\psi}_0\bigr>-\bigl(\hat{O}^{-1}\hat{Q}^{(1)}e^T+\hat{O}^{-1}\hat{Q}^{(2)}e^{2T}\bigr)\bigl\vert\boldsymbol{\psi}\bigr>.
\end{eqnarray}

\noindent
Equation (\ref{THEQUANTUM1}) can be rearranged into the form

\begin{eqnarray}
\label{THEQUANTUM2}
\bigl(1+\hat{q}_1+\hat{q}_2\bigr)\bigl\vert\boldsymbol{\psi}\bigr>=\bigl\vert\boldsymbol{\psi}_0\bigr>,
\end{eqnarray}

\noindent
where we have defined

\begin{eqnarray}
\label{THEQUANTUM3}
\hat{q}_1\equiv\hat{O}^{-1}\hat{Q}^{(1)};~~\hat{q}_2\equiv\hat{O}^{-1}\hat{Q}^{(2)}.
\end{eqnarray}

\noindent
Then (\ref{THEQUANTUM2}) can be rearranged into the form

\begin{eqnarray}
\label{THEQUANTUM4}
\bigl\vert\boldsymbol{\psi}\bigr>=\Bigl({1 \over {1+\hat{q}_1+\hat{q}_2}}\Bigr)\bigl\vert\boldsymbol{\psi}_0\bigr>.
\end{eqnarray}

\noindent
For labelling purposes let $\bigl\vert\lambda\bigr>\in{Ker}\{\hat{O}\}$.  The action of the individual operators is are given by

\begin{eqnarray}
\label{LEGION9}
\hat{q}_1\bigl\vert\lambda\bigr>=\hat{O}^{-1}\hat{Q}^{(1)}e^T\bigl\vert\lambda\bigr>
=\hat{O}^{-1}\hat{Q}^{(1)}\bigl\vert\lambda+\mu^{\prime}\bigr>
=E^{(1)}_{\lambda+\mu^{\prime}}(\alpha,\beta)\bigl\vert\lambda+\mu^{\prime}\bigr>,
\end{eqnarray}

\noindent
and for $\hat{q}_2$ by

\begin{eqnarray}
\label{LEGION10}
\hat{q}_2\bigl\vert\lambda\bigr>=\hat{O}^{-1}\hat{Q}^{(2)}e^{2T}\bigl\vert\lambda\bigr>
=\hat{O}^{-1}\hat{Q}^{(2)}\bigl\vert\lambda+2\mu^{\prime}\bigr>
=E^{(2)}_{\lambda+2\mu^{\prime}}(\alpha,\beta)\bigl\vert\lambda+2\mu^{\prime}\bigr>,
\end{eqnarray}

\noindent
where

\begin{eqnarray}
\label{WHERE}
E^{(1)}_{\lambda+k\mu^{\prime}}(\alpha,\beta)={{\lambda+k\mu^{\prime}+{1 \over 3}(\alpha+\beta)} \over {(\lambda+k\mu^{\prime})(\lambda+k\mu^{\prime}+\alpha)(\lambda+k\mu^{\prime}+\beta)}};\nonumber\\
E^{(2)}_{\lambda+k\mu^{\prime}}(\alpha,\beta)={{(\lambda+k\mu^{\prime}+\gamma^{-}(\alpha,\beta))
(\lambda+k\mu^{\prime}+\gamma^{+}(\alpha,\beta))} \over {(\lambda+k\mu^{\prime})(\lambda+k\mu^{\prime}+\alpha)(\lambda+k\mu^{\prime}+\beta)}}.
\end{eqnarray}

\noindent
Then the full solution to (\ref{THEQUANTUM4}) can be written as

\begin{eqnarray}
\label{LEGION12}
\bigl\vert\boldsymbol{\psi}\bigr>=\Bigl(\sum_{n=0}^{\infty}(-1)^n(\hat{q}_1+\hat{q}_2)^n\Bigr)\bigl\vert\lambda\bigr>
=\sum_{n=1}^{\infty}(-1)^n\hat{q}_{k_n}\hat{q}_{k_{n-1}}\dots\hat{q}_{k_2}\hat{q}_{k_1}\bigl\vert\lambda\bigr>.
\end{eqnarray}

\noindent
At each order $n$ there are $2^n$ terms in the expansion, and the summation must be made over all permutations of indices with the operator ordering preserved.  The indices $k_n$ take on the value of $1$ or $2$.  To illustrate for the first two terms, starting with the first term we have

\begin{eqnarray}
\label{LEGION13}
\hat{q}_{k_1}\bigl\vert\lambda\bigr>=E^{(k_1)}_{\lambda+k_1\mu^{\prime}}\bigl\vert\lambda+k_1\mu^{\prime}\bigr>,
\end{eqnarray}

\noindent
where we have suppressed the $(\alpha,\beta)$ labels to avoid cluttering up the notation.  For the second term we have

\begin{eqnarray}
\label{LEGION14}
\hat{q}_{k_2}\hat{q}_{k_1}\bigl\vert\lambda\bigr>=E^{(k_1)}_{\lambda+k_1\mu^{\prime}}E^{(k_2)}_{\lambda+(k_1+k_2)\mu^{\prime}}
\bigl\vert\lambda+(k_1+k_2)\mu^{\prime}\bigr>.
\end{eqnarray}

\noindent
The $N^{th}$ term is given by

\begin{eqnarray}
\label{LEGION15}
\prod_{n=1}^N\hat{q}_{k_n}\bigl\vert\lambda\bigr>=\prod_{n=1}^NE^{(k_n)}_{\lambda+(k_1+k_2+\dots{k}_n)\mu^{\prime}}\bigl\vert\lambda+(k_1+k_2+\dots{k}_n)\mu^{\prime}\bigr>
\end{eqnarray}

\noindent
Our main concern is the convergence of the full series (\ref{LEGION12}), which we will show using norm inequalities.  For large $k$, equation (\ref{WHERE}) implies that

\begin{eqnarray}
\label{LEGION16}
\vert{E}^{(k_1)}_{\lambda+k\mu^{\prime}}\vert\leq{1 \over {k\mu^{\prime}}}.
\end{eqnarray}

\noindent
So for large $N$, the labels $\alpha$ and $\beta$ become unimportant and each term in the product in (\ref{LEGION15}) satisfies the following bound

\begin{eqnarray}
\label{LEGION17}
\bigl\vert{E}^{k_n}_{\lambda+(k_1+k_2+\dots{k}_N)\mu^{\prime}}\bigr\vert\leq{1 \over {N\mu^{\prime}}}.
\end{eqnarray}

\noindent
But there are $2^N$ terms, corresponding to the different permutations of the indices $k_1k_2\dots{k}_N$.  So the full series is bounded by

\begin{eqnarray}
\label{LEGION18}
\sum_N\Bigl(\prod_{n=1}^N\hat{q}_{k_n}\Bigr)\bigl\vert\lambda\bigr>\leq\sum_N{{2^N} \over {N!}}\Bigl({{e^{2\vert{T}\vert}} \over {\mu^{\prime}}}\Bigr)^N
=\hbox{exp}\Bigl({{2e^{2\vert{T}\vert}} \over {\mu^{\prime}}}\Bigr),
\end{eqnarray}

\noindent
which is a convergent function.  The solution to the Hamiltonian constraint converges, since each term of (\ref{LEGION18}) vanishes as $N\rightarrow\infty$, $\forall{T}$.

\section{Conclusion}

\noindent
This paper provides a first step toward the realization of Penrose's idea of the nonlinear graviton.  We have demonstrated the existence of gravitons, both at the linearized and at the non linearized level, using a new description of nonmetric complex general relativity defined on the phase space $\Omega_{Inst}=(\widetilde{\sigma}^i_a,A^a_i)$.  The conventional method in the linearization of gravity is to expand the spacetime metric in fluctuations about a fixed background, typically a flat Minkowski spacetime.  In this paper we have chosen for a background spacetimes whose semiclassical orbits arise from the Kodama state $\boldsymbol{\psi}_{Kod}$.  We have expanded the theory 
on $\Omega_{Inst}$ in fluctuations about $\boldsymbol{\psi}_{Kod}$, constructing the Hilbert space of states annihilated by the constraints both at the linearized and at the nonlinearized level.  In each case there were two physical degrees of freedom per point, implying the preservation of these degrees of freedom under linearization.  Additionally, it is hoped that the results of this paper have provided an addressal of the issues surrounding the Kodama 
state $\boldsymbol{\psi}_{Kod}$ raised in \cite{WITTEN1} and in \cite{NORMKOD}.  Clearly, the Kodama state provides a natural background for quantizable flucutations when one uses the phase space variables on $\Omega_{Inst}$.  It is clear also that when restricted to the diagonal connection $A^a_i=\delta^a_iA^a_a$ used in this paper, that the integrand of the Kodama state is proportional to $(\hbox{det}A)=A^1_1A^2_2A^3_3$, which plays the role of a time variable on configuration space.  The addressal of the issue of the normalizabilty of $\boldsymbol{\psi}_{Kod}$ then is simply that one does not normalize a wavefunction in time.  However, one should normalize wavefunctions with respect to the physical degrees of freedom orthogonal to the time direction.  And we have done so using a Gaussian measure for normalization, which ensures square integrability of the wavefunctions on the gravitational Hilbert space.  The well-definedness of these wavefunctions with respect to their time dependence arises due to the convergence of the infinite series constituting the solution to the quantum Hamiltonian constraint.  The notion of the Chern--Simons functional as a time variable on configuration space has been proposed in \cite{SOO} and \cite{SOO1}.

\end{document}